\pdfoutput = 1
\documentclass[article,preprint,superscriptaddress]{revtex4-1}
\usepackage{graphicx}
\usepackage{dcolumn}
\usepackage{amssymb}
\usepackage{amsmath}
\usepackage{bm}
\usepackage{pifont}
\usepackage{epsfig}
\usepackage{psfrag}
\usepackage[usenames]{xcolor}
\begin{document}

\title{Dynamical facilitation governs glassy dynamics in suspensions of colloidal ellipsoids}
\author{Chandan K. Mishra}
\affiliation{Chemistry and Physics of Materials Unit, Jawaharlal Nehru Centre for Advanced Scientific Research, Jakkur, Bangalore - 560064, INDIA}
\author{K. Hima Nagamanasa}
\affiliation{Chemistry and Physics of Materials Unit, Jawaharlal Nehru Centre for Advanced Scientific Research, Jakkur, Bangalore - 560064, INDIA}
\author{Rajesh Ganapathy}
\affiliation{International Centre for Materials Science, Jawaharlal Nehru Centre for Advanced Scientific Research, Jakkur, Bangalore - 560064, INDIA}
\author{A. K. Sood}
\affiliation{International Centre for Materials Science, Jawaharlal Nehru Centre for Advanced Scientific Research, Jakkur, Bangalore - 560064, INDIA}
\affiliation{Department of Physics, Indian Institute of Science, Bangalore - 560012, INDIA}
\author{Shreyas Gokhale$^{\ast}$}
\affiliation{Department of Physics, Indian Institute of Science, Bangalore - 560012, INDIA}


%
\begin{abstract}
{One of the greatest challenges in contemporary condensed matter physics is to ascertain whether the formation of glasses from liquids is fundamentally thermodynamic or dynamic in origin. While the thermodynamic paradigm has dominated theoretical research for decades, the purely kinetic perspective of the dynamical facilitation (DF) theory has attained prominence in recent times. In particular, recent experiments and simulations have highlighted the importance of facilitation using simple model systems composed of spherical particles. However, an overwhelming majority of liquids possess anisotropy in particle shape and interactions and it is therefore imperative to examine facilitation in complex glass-formers. Here, we apply the DF theory to systems with orientational degrees of freedom as well as anisotropic attractive interactions. By analyzing data from experiments on colloidal ellipsoids, we show that facilitation plays a pivotal role in translational as well as orientational relaxation. Further, we demonstrate that the introduction of attractive interactions leads to spatial decoupling of translational and rotational facilitation, which subsequently results in the decoupling of dynamical heterogeneities. Most strikingly, the DF theory can predict the existence of reentrant glass transitions based on the statistics of localized dynamical events, called excitations, whose duration is substantially smaller than the structural relaxation time. Our findings pave the way for systematically testing the DF approach in complex glass-formers and also establish the significance of facilitation in governing structural relaxation in supercooled liquids.}
\end{abstract}

\date{\today}
\draft
\maketitle
\renewcommand{\thefootnote}



The transformation of liquids into glasses is as ubiquitous as it is enigmatic. From the formation of obsidian during volcanic eruptions \cite{shackley2005obsidian} and fabrication of superstrong metallic glasses \cite{demetriou2011damage} to exotic forms of slow dynamics in crystals of colloidal dimers \cite{gerbode2010glassy} and Janus particles \cite{jiang2014orientationally}, glass formation pervades nature, industry and academia. A vast majority of molecular glass-forming materials exhibit anisotropy in shape and inter-particle interactions, which often has a profound influence on their glassy dynamics. The rapidly expanding repertoire of chemists has made it possible to design colloidal particles of desired shape and interactions that can serve as realistic experimental analogues of these molecular liquids \cite{glotzer2007anisotropy}. By contrast, prominent theories like the Adam-Gibbs theory \cite{adam1965temperature}, Random First-Order Transition (RFOT) theory \cite{kirkpatrick1989scaling,lubchenko2007theory} and the Dynamical Facilitation (DF)theory \cite{garrahan2002geometrical,chandler2009dynamics} have been tested predominantly on spherical glass-formers with isotropic interactions, which exhibit gross features of glassy dynamics, but fail to capture the nuances of vitrification in complex systems. 

The discovery of growing static \cite{biroli2008thermodynamic,karmakar2009growing,tanaka2010critical,kurchan2011order,karmakar2012direct,dunleavy2012using} and dynamic \cite{donati1998stringlike,weeks2000three,kegel2000direct,berthier2005direct,kob2012non} length scales appears to support the thermodynamic perspective of the Adam-Gibbs and RFOT theories. However, the growth in static length scales over the dynamical range accessible to numerical simulations is often minuscule and much smaller than the corresponding growth in dynamic length scales \cite{kob2012non,charbonneau2013decorrelation}. This renders any causal connection between growing static length scales and growing timescales doubtful \cite{charbonneau2013decorrelation}. Moreover, recent simulations \cite{keys2011excitations} and colloid experiments \cite{gokhale2014growing} have shown that growing dynamical correlations in the form of string-like cooperative motion emerge naturally within the purely kinetic approach of the DF theory. To compound matters further, facilitation is present even within the RFOT framework, albeit as a consequence of slow dynamics rather than a cause \cite{bhattacharyya2008facilitation}. Thus, while DF has been shown to exist \cite{keys2011excitations,gokhale2014growing,vogel2004spatially,bergroth2005examination,candelier2009building,candelier2010spatiotemporal}, its relative importance as a mechanism of structural relaxation is still debated \cite{berthier2011theoretical,candelier2010dynamical,biroli2013perspective}. The application of the DF approach to complex glass-formers will therefore not only enhance our understanding of glass transitions in these systems, but also help ascertain the relevance of facilitation in governing structural relaxation. 

Here, we apply the DF theory to elucidate glass formation in suspensions of colloidal ellipsoids with repulsive as well as attractive interactions. The DF theory claims that structural relaxation in glass-forming liquids proceeds via a process known as dynamical facilitation, whereby localized mobile regions, termed excitations, mediate motion in neighboring regions in a manner that conserves mobility \cite{garrahan2002geometrical,chandler2009dynamics}. We first show that the notions of localized excitations and facilitated dynamics can be extended even to orientational relaxation. Next, we demonstrate that the spatial decoupling of dynamical heterogeneities observed in colloid experiments stems from the spatial decoupling of rotational and translational facilitation. Most importantly, the DF theory can predict the existence of recently observed reentrant glass transitions \cite{mishra2013two} from the density dependence of the concentration of excitations. Our findings not only highlight the importance of facilitated dynamics in anisotropic glass-formers, but also reinforce the claim that in the broader context of the glass transition, facilitation dominates structural relaxation.   

\section*{Results and Discussion}
\subsection*{Facilitated dynamics of rotational and translational excitations}
\begin{figure}[tbp]
\includegraphics[width=0.5\textwidth]{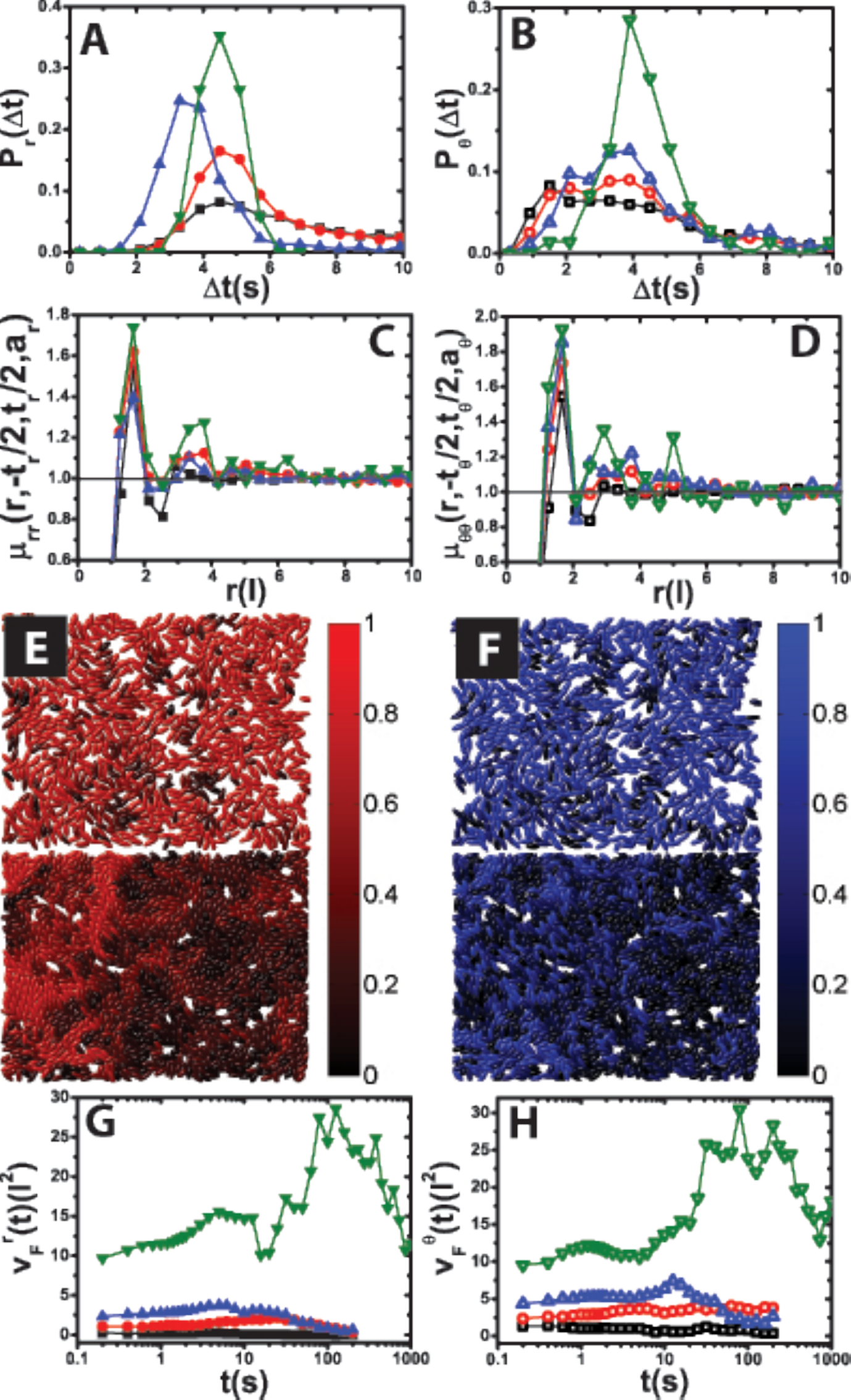}
\caption{Facilitation in translational and orientational degrees of freedom. (A) Instanton time distribution $P_r(\Delta t)$ for translational excitations of size $a_r = 0.33l$ for $\phi =$ 0.68 ({\color{black} $\boldsymbol \blacksquare$}), $\phi =$ 0.73 ({\color{red} $\boldsymbol \bullet$}), $\phi =$ 0.76 ({\color{blue} $\boldsymbol \blacktriangle$}) and $\phi =$ 0.79 ({\color{green!50!black} $\boldsymbol \blacktriangledown$}). (B) Instanton time distribution $P_{\theta}(\Delta t)$ for rotational excitations of size $a_{\theta} = 10^{\circ}$ for $\phi =$ 0.68 ({\color{black} $\boldsymbol \square$}), $\phi =$ 0.73 ({\color{red} $\boldsymbol \circ$}), $\phi =$ 0.76 ({\color{blue} $\boldsymbol \triangle$}) and $\phi =$ 0.79 ({\color{green!50!black} $\boldsymbol \triangledown$}). (C) The function $\mu_{rr}(r,-t_r/2,t_r/2;a_r)$ for $a_r = 0.33l$ for various $\phi$s. The colors and symbols are identical to those in (A). (D) The function $\mu_{\theta\theta}(r,-t_{\theta}/2,t_{\theta}/2;a_{\theta})$ for $a_{\theta} = 10^{\circ}$ for various $\phi$s. The colors and symbols are identical to those in (B). (E) The translational displacement field $|\Delta \textbf{r}_i(t_r)|$ normalized by $a_r = 0.33l$.(F) The rotational displacement field $|\Delta \theta_i(t_{\theta})|$ normalized by $a_\theta = 10^{\circ}$. In (E) and (F), the top panels correspond to $\phi =$0.73 and the bottom panels correspond to $\phi =$0.79. (G) Facilitation volume for translational excitations, $v_{F}^{r}(t)$, for $a_r = 0.33l$ for various $\phi$s. The colors and symbols are identical to those in (A). (H) Facilitation volume for rotational excitations, $v_{F}^{\theta}(t)$, for $a_\theta = 10^{\circ}$ for various $\phi$s. The colors and symbols are identical to those in (B).}
\label{Fig1}
\end{figure}

We analysed data from video microscopy experiments \cite{mishra2013two} on quasi-2D monolayers of colloidal polystyrene ellipsoids of aspect ratio $\alpha =$ 2.1, with semi-major axis $l = 2.1$ $\mu$m and semi-minor axis $w = 1$ $\mu$m (See Materials and Methods). We first identified translational and rotational excitations for ellipsoids with purely repulsive interactions by following the prescription of \cite{keys2011excitations}. Accordingly, a particle was said to be associated with a translational (rotational) excitation of size $a_r$ ($a_{\theta}$) and `instanton' time $\Delta t_r$ ($\Delta t_{\theta}$), if it underwent a linear (angular) displacement of magnitude $a_r$ ($a_{\theta}$) over a time interval $\Delta t_r$ ($\Delta t_{\theta}$), and persisted in its initial and final positions for at least as long as the instanton time duration. To identify excitations, we first generated coarse-grained translational and rotational trajectories, $\bar{\textbf{r}}_{i}(t)$ and $\bar{\theta}_{i}(t)$, respectively, for every particle $i$, and computed the functionals
\begin{eqnarray}
h_{i}^{r}(t,t_{r};a_r) = \prod\limits_{t' = t_{r}/2 - \Delta t_r}^{t_{r}/2} \mathcal{H}(|\bar{\textbf{r}}_{i}(t+t')-\bar{\textbf{r}}_{i}(t-t')| - a_r)\nonumber
\\
h_{i}^{\theta}(t,t_{\theta};a_{\theta}) = \prod\limits_{t' = t_{\theta}/2 - \Delta t_{\theta}}^{t_{\theta}/2} \mathcal{H}(|\bar{\theta}_{i}(t+t')-\bar{\theta}_{i}(t-t')| - a_{\theta})
\end{eqnarray}
Here, $\mathcal{H}(x)$ is the Heaviside step function and $t_r$ and $t_{\theta}$ are `commitment' times that are typically chosen to be $\sim$ 3-4 times the respective mean instanton times \cite{gokhale2014growing}. In this work, we varied $a_r$ and $a_{\theta}$ in the range $0.33l \leq a_r \leq 0.83l$ and $10^{\circ} \leq a_{\theta} \leq 25^{\circ}$, respectively. Analogous to spherical colloids \cite{gokhale2014growing}, the distributions of instanton times for translational (P$_{r}$($\Delta t$)) as well as rotational (P$_{\theta}$($\Delta t$)) excitations remain fairly localized for all area fractions $\phi$ (Fig. \ref{Fig1}A-B). Further, the peak instanton times $\Delta t_p^{r}$ and $\Delta t_p^{\theta}$ do not grow with $\phi$ and are much smaller than the corresponding structural relaxation times $\tau_{\alpha}$ at large $\phi$ \cite{mishra2013two}. Thus, rotational excitations can not only be defined using the procedure developed in \cite{keys2011excitations}, but are also temporally localized, as postulated by the DF theory.

Next, we quantified the spatial extent of rotational and translational excitations, by computing the functions
\begin{eqnarray}
\label{MuFunctions}
\nonumber\mu_{rr}(r,t,t';a_r) = \frac{1}{\rho\mu_{\infty}^{r}(t'-t)\langle h_{1}^{r}(0,t_{r};a_r) \rangle} \Bigg\langle h_{1}^{r}(0,t_{r};a_r)\sum\limits_{i \neq 1}^{N}|\bar{\textbf{r}}_{i}(t')-\bar{\textbf{r}}_{i}(t)|\delta(\bar{\textbf{r}}_{i}(t) - \bar{\textbf{r}}_{1}(t) - \bar{\textbf{r}}) \Bigg\rangle
\\
\nonumber \mu_{\theta\theta}(r,t,t';a_{\theta}) = \frac{1}{\rho\mu_{\infty}^{\theta}(t'-t)\langle h_{1}^{\theta}(0,t_{\theta};a_{\theta}) \rangle} \Bigg\langle h_{1}^{\theta}(0,t_{\theta};a_{\theta})\sum\limits_{i \neq 1}^{N}|\bar{\theta}_{i}(t')-\bar{\theta}_{i}(t)|\delta(\bar{\textbf{r}}_{i}(t) - \bar{\textbf{r}}_{1}(t) - \bar{\textbf{r}}) \Bigg\rangle
\\
\end{eqnarray}
Here, $\rho$ is the particle number density, $\mu_{\infty}^{r}(t) = \langle|\bar{\textbf{r}}_{i}(t) - \bar{\textbf{r}}_{i}(0)|\rangle$ and $\mu_{\infty}^{\theta}(t) = \langle|\bar{\theta}_{i}(t) - \bar{\theta}_{i}(0)|\rangle$. The functions $\mu_{rr}(r,-t_r/2,t_r/2;a_r)$ and $\mu_{\theta\theta}(r,-t_{\theta}/2,t_{\theta}/2;a_{\theta})$ respectively yield the translational and rotational displacement density at a distance $r$ from a translational or rotational excitation of a given size centred at the origin at time $t=0$, over the corresponding commitment times $t_r$ and $t_{\theta}$. We observed that for all $\phi$, $\mu_{rr}(r,-t_r/2,t_r/2;a_r)$ and $\mu_{\theta\theta}(r,-t_{\theta}/2,t_{\theta}/2;a_{\theta})$ stabilize at 1 within 6-8$l$ (Fig. \ref{Fig1}C-D), confirming that in accordance with the DF theory, rotational as well as translational excitations are spatially localized objects that do not exhibit any growth as the glass transition is approached. 

One of the two central tenets of the DF theory is a decrease in the concentration of excitations on approaching the glass transition. This concentration can be visualized from the normalized translational and rotational displacement fields $|\Delta \textbf{r}_i(t_r)|/a_r$ and $|\Delta \theta_i(t_{\theta})|/a_{\theta}$, respectively. Figure \ref{Fig1}E-F shows these displacement fields for $\phi =$ 0.73 and $\phi =$ 0.79. In Fig. \ref{Fig1}E, the red regions denote particles that undergo displacements larger than the excitation size $a_r = 0.33l$ over the corresponding commitment time $t_r$, and therefore indicate the presence of translational excitations. Similarly, the blue regions in Fig. \ref{Fig1}F denote particles that undergo angular displacements larger than $a_{\theta} = 10^{\circ}$ over $t_{\theta}$, thus serving as an indicator of the concentration of rotational excitations. It is obvious from Fig. \ref{Fig1}E-F that the concentration of translational as well as rotational excitations indeed decreases with $\phi$, in concord with the DF theory. The concentration of translational excitations, $c_r$, and rotational excitations, $c_{\theta}$, can be formally defined as
\begin{eqnarray}
\label{ExcitConc}
\nonumber c_{r} = \Bigg\langle \frac{1}{Vt_{r}} \sum\limits_{i=1}^{N} h_{i}^{r}(0,t_{r};a_r)\Bigg\rangle \\
c_{\theta} = \Bigg\langle \frac{1}{Vt_{\theta}} \sum\limits_{i=1}^{N} h_{i}^{\theta}(0,t_{\theta};a_{\theta})\Bigg\rangle
\end{eqnarray}  
where $V$ is the volume and $N$ is the total number of particles. $c_r$ and $c_{\theta}$ play an especially important role in the prediction of reentrant glass transitions in translational and rotational degrees of freedom.

The second major claim of the DF theory is that mobility cannot be spontaneously created or destroyed, and hence excitations, which are carriers of mobility, must always occur in the vicinity of existing excitations. From the functions defined in Eqn. \eqref{MuFunctions} we extract facilitation volumes, which serve as measures of facilitated dynamics in translational as well as rotational degrees of freedom. 
\begin{eqnarray}
\label{FacVolume}
\nonumber v_{F}^{r}(t) = \int\Bigg[\frac{\mu_{rr}(r,t_r/2,t;a_r)}{g(r)} - 1\Bigg]d\textbf{r} \\
v_{F}^{\theta}(t) = \int\Bigg[\frac{\mu_{\theta\theta}(r,t_{\theta}/2,t;a_{\theta})}{g(r)} - 1\Bigg]d\textbf{r}
\end{eqnarray}
Here, $g(r)$ is the radial pair correlation function. These facilitation volumes quantify the size of the region around an excitation in which structural relaxation at time $t$ can be unambiguously attributed to the presence of the initial excitation. We observe once again, that in close analogy with colloidal spheres \cite{gokhale2014growing}, the profiles of $v_{F}^{r}(t)$ as well as $v_{F}^{\theta}(t)$ generically exhibit a maximum (Fig. \ref{Fig1}G-H). In general, facilitation volumes depend on the concentration of excitations, which in turn depends on the excitation size \cite{keys2011excitations}. Here, we have chosen $a_r$ and $a_{\theta}$ such that the peak instanton times for translational and rotational excitations are comparable, i.e. $\Delta t_p^{r} \sim \Delta t_p^{\theta}$ (Fig. \ref{Fig1}A-B), a fact that is also reflected in the facilitation volumes (Fig. \ref{Fig1}G-H). Further, we see the maximum values $v_{F}^{r}(t)$ and $v_{F}^{\theta}(t)$ as well as the time taken to reach the maximum increase with $\phi$, showing that facilitated dynamics becomes increasingly apparent on approaching the glass transition \cite{keys2011excitations,elmatad2012manifestations}, even in systems with orientational degrees of freedom (See Video S1 and Video S2 for a visualization of facilitated dynamics in translational and rotational degrees of freedom, respectively). 

\subsection*{Coupling between rotational and translational facilitation}
\begin{figure}[tbp]
\includegraphics[width=0.7\textwidth]{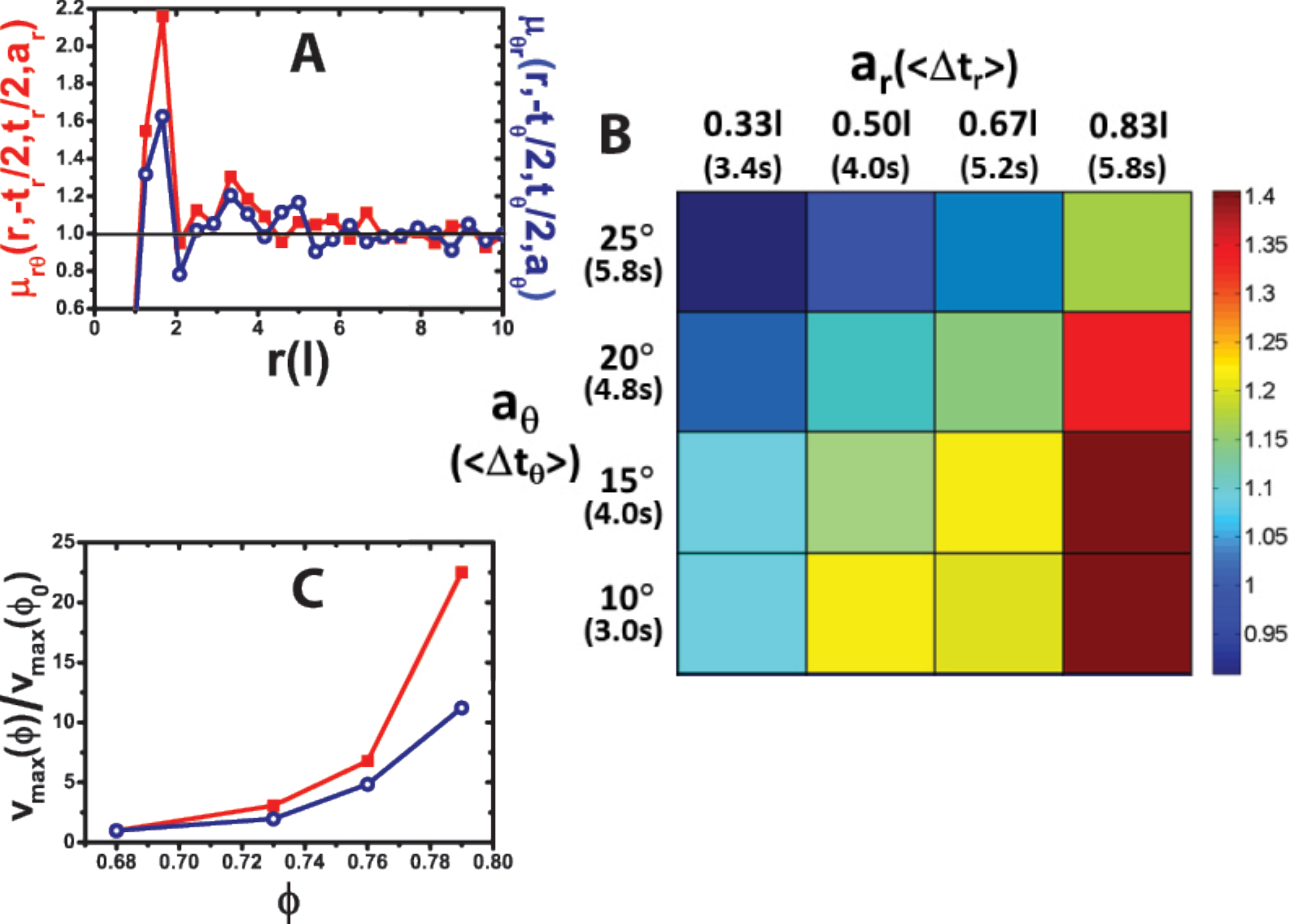}
\caption{Coupling between translational and rotational facilitation. (A) The off-diagonal functions $\mu_{r\theta}(r,-t_r/2,t_r/2;a_r)$ ({\color{red} $\boldsymbol \blacksquare$}) and $\mu_{\theta r}(r,-t_{\theta}/2,t_{\theta}/2;a_{\theta})$ ({\color{blue} $\boldsymbol \circ$}) for $\phi =$0.76. The values $a_r = 0.5l$ and $a_{\theta} = 15^{\circ}$ are chosen such that $\langle \Delta t\rangle_r \sim \langle \Delta t\rangle_{\theta}$. (B) Facilitation volume profiles for the off-diagonal functions in (C). (B) The asymmetry parameter $F(a_r,a_{\theta},t_m)$ for various combinations of $a_r$ and $a_{\theta}$. The corresponding values of $\langle \Delta t\rangle_r$ and $\langle \Delta t\rangle$ are enclosed in parentheses. (C) The maxima of the off-diagonal facilitation volumes, $v_{\text{max}}^{r}$ and $v_{\text{max}}^{\theta}$ as a function of $\phi$, normalized by their respective values at $\phi_0 =$ 0.68. The colors and symbols are identical to those in (A).}
\label{Fig2}
\end{figure}

Since excitations are, by definition, carriers of mobility, it is natural to identify them from particle displacements in systems with purely translational degrees of freedom, such as those in \cite{keys2011excitations,gokhale2014growing}. In ellipsoids, however, mobility is distributed between particle translations and rotations, and hence, facilitation in the two degrees of freedom does not occur independently. In particular, translational excitations can facilitate rotational relaxation and vice versa. To quantify this coupling in translational and rotational facilitation, we defined mixed variants of the functions in Eqn. \eqref{MuFunctions}.
\begin{eqnarray}
\label{MixedMu}
\nonumber\mu_{r\theta}(r,t,t';a_r) = \frac{1}{\rho\mu_{\infty}^{r}(t'-t)\langle h_{1}^{r}(0,t_{r};a_r) \rangle} \Bigg\langle h_{1}^{r}(0,t_{r};a_r)\sum\limits_{i \neq 1}^{N}|\bar{\theta}_{i}(t')-\bar{\theta}_{i}(t)|\delta(\bar{\textbf{r}}_{i}(t) - \bar{\textbf{r}}_{1}(t) - \bar{\textbf{r}}) \Bigg\rangle
\\
\nonumber \mu_{\theta r}(r,t,t';a_{\theta}) = \frac{1}{\rho\mu_{\infty}^{\theta}(t'-t)\langle h_{1}^{\theta}(0,t_{\theta};a_{\theta}) \rangle} \Bigg\langle h_{1}^{\theta}(0,t_{\theta};a_{\theta})\sum\limits_{i \neq 1}^{N}|\bar{\textbf{r}}_{i}(t')-\bar{\textbf{r}}_{i}(t)|\delta(\bar{\textbf{r}}_{i}(t) - \bar{\textbf{r}}_{1}(t) - \bar{\textbf{r}}) \Bigg\rangle
\\
\end{eqnarray}
The mixed, or `off-diagonal' functions $\mu_{r\theta}(r,t,t';a_r)$ and $\mu_{\theta r}(r,t,t';a_{\theta})$ exhibit qualitatively similar behavior to their pure, or `diagonal' counterparts (Fig. \ref{Fig2}A). Interestingly, we see that the first peak of $\mu_{r\theta}(r,t,t';a_r)$ is sharper than that of $\mu_{\theta r}(r,t,t';a_{\theta})$, which suggests that the influence of translational excitations on rotational dynamics is stronger than that of rotational excitations on translational dynamics. To investigate whether this imbalance in facilitation is a generic feature of our system, we calculated the asymmetry parameter $F(a_r,a_{\theta},t_m) = \mu_{r\theta}^{max}(t_m) / \mu_{\theta r}^{max}(t_m)$ for 16 different combinations of $a_r$ and $a_{\theta}$ for $\phi =$ 0.73. Here, $\mu_{r\theta}^{max}(t_m)$ and $\mu_{\theta r}^{max}(t_m)$ are the first maxima of $\mu_{r\theta}(r,-t_m/2,t_m/2;a_r)$ and $\mu_{\theta r}(r,-t_m/2,t_m/2;a_{\theta})$, respectively, and $t_m = \text{max}(t_r,t_{\theta})$. We find that $F(a_r,a_{\theta},t_m)$ is greater than 1 for 14 of the 16 combinations of $a_r$ and $a_{\theta}$ considered (Fig. \ref{Fig2}B). For the remaining two combinations, we find that $\Delta t_p^{\theta} > \Delta t_p^{r}$, suggesting that the low values of $F(a_r,a_{\theta},t_m)$ in these cases stem from the disparity in the sizes of rotational and translational excitations. For cases in which $\Delta t_p^{r} \sim \Delta t_p^{\theta}$, $F(a_r,a_{\theta},t_m) \sim 1.2$, which shows that the impact of translational excitations on rotational relaxation is indeed more profound as compared to that of rotational ones on translational relaxation. From the nature of collisions in a dense, orientationally and translationally disordered system, one would expect a large translation of a particle to induce large rotations in its vicinity in addition to inducing translations. Further large rotational motions would be likelier to induce rotational rather than translational motion in the particle's neighborhood. The observed asymmetry in facilitated dynamics is consistent with this intuitive picture. Next, we computed the corresponding off-diagonal facilitation volumes $v_F^{r\theta}(t)$ and $v_F^{\theta r}(t)$, defined analogously to Eqn. \ref{FacVolume} (Fig. S1), and found that the maximum of $v_F^{r\theta}(t)$ increases more rapidly with $\phi$ as compared to that of $v_F^{\theta r}(t)$ (Fig. \ref{Fig2}C). This implies that the asymmetry in facilitation becomes increasingly pronounced on approaching the glass transition. 

\subsection*{Spatial decoupling of dynamical heterogeneities and prediction of reentrant glass transitions} 
\begin{figure}[tbp]
\includegraphics[width=0.6\textwidth]{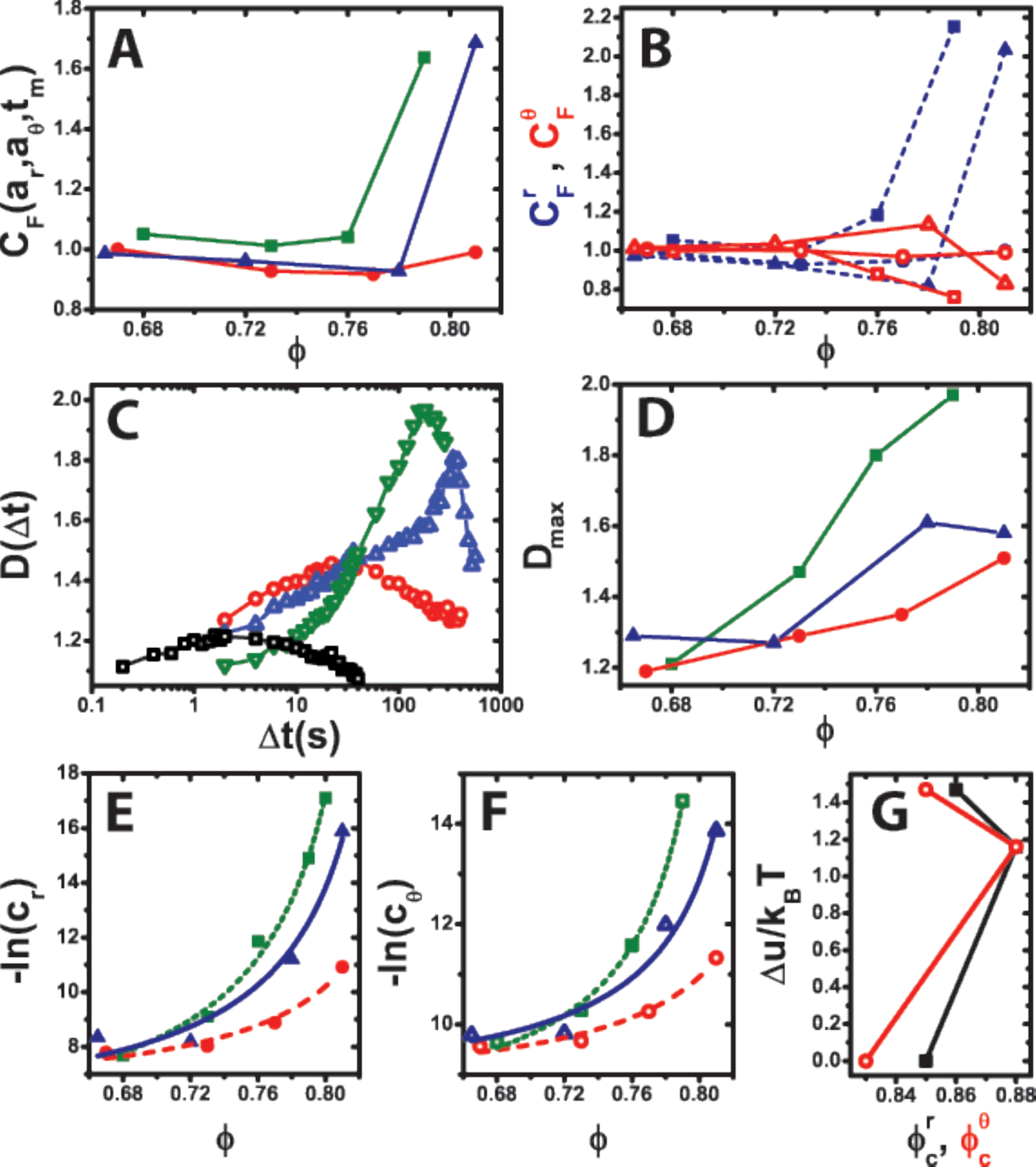}
\caption{Spatial decoupling of heterogeneities and prediction of reentrant glass transitions. (A) The coupling coefficient for facilitation, $C_{F}(a_r,a_{\theta},t_m)$, with $a_r = 0.5l$ and $a_{\theta} = 20^{\circ}$, as a function of $\phi$ for $\Delta u/k_BT =$ 0 ({\color{green!50!black} $\boldsymbol \blacksquare$}), $\Delta u/k_BT =$ 1.16 ({\color{red} $\boldsymbol \bullet$}) and $\Delta u/k_BT =$ 1.47 ({\color{blue!60!black} $\boldsymbol \blacktriangle$}). (B) The translational and rotational contributions to $C_{F}(a_r,a_{\theta},t_m)$, $C_{F}^{r}$ (filled blue symbols) and $C_{F}^{\theta}$ (hollow red symbols), respectively, as a function of $\phi$ for $\Delta u/k_BT =$ 0 (squares), $\Delta u/k_BT =$ 1.16 (circles) and $\Delta u/k_BT =$ 1.47 (triangles). (C) The coupling function for dynamical heterogeneities, $D(\Delta t)$, in the purely repulsive case, $\Delta u/k_BT =$ 0, for $\phi =$ 0.68 ({\color{black} $\boldsymbol \square$}), $\phi =$ 0.73 ({\color{red} $\boldsymbol \circ$}), $\phi =$ 0.76 ({\color{blue} $\boldsymbol \triangle$}) and $\phi =$ 0.79 ({\color{green!50!black} $\boldsymbol \triangledown$}). (D) The coupling coefficient for dynamical heterogeneities, $D_{\text{max}}$, as a function of $\phi$ for $\Delta u/k_BT =$ 0 ({\color{green!50!black} $\boldsymbol \blacksquare$}), $\Delta u/k_BT =$ 1.16 ({\color{red} $\boldsymbol \bullet$}) and $\Delta u/k_BT =$ 1.47 ({\color{blue!60!black} $\boldsymbol \blacktriangle$}). (E) The $\phi$ dependence of the concentration of translational excitations $c_r$ for $a_r = 0.5l$, for $\Delta u/k_BT =$ 0 ({\color{green!50!black} $\boldsymbol \blacksquare$}), $\Delta u/k_BT =$ 1.16 ({\color{red} $\boldsymbol \bullet$}) and $\Delta u/k_BT =$ 1.47 ({\color{blue!60!black} $\boldsymbol \blacktriangle$}). (F) The $\phi$ dependence of the concentration of rotational excitations $c_{\theta}$ for $a_{\theta} = 20^{\circ}$ for $\Delta u/k_BT =$ 0 ({\color{green!50!black} $\boldsymbol \square$}), $\Delta u/k_BT =$ 1.16 ({\color{red} $\boldsymbol \circ$}) and $\Delta u/k_BT =$ 1.47 ({\color{blue!60!black} $\boldsymbol \triangle$}). In (E) and (F), the curves are empirical fits of the form $\phi_0 + A(\phi_c - \phi)^{-1}$. (G) The translational glass transition $\phi_{c}^{r}$ ({\color{black} $\boldsymbol \blacksquare$}) and rotational glass transition $\phi_{c}^{\theta}$ ({\color{red} $\boldsymbol \circ$}) obtained from fits to the curves in (E) and (F), for various values of $\Delta u/k_{B}T$.}
\label{Fig3}
\end{figure}

Having elucidated the nature of facilitated dynamics in the case of ellipsoids with short-ranged repulsions, we turn our attention to the influence of attractive depletion interactions on glass formation. Depletion interactions between particles are known to depend on local curvature \cite{asakura2004interaction,schiller2011interactions}. For ellipsoids, therefore, these interactions are anisotropic and favour the alignment of particles along their major axis \cite{mishra2013two}. We first examine the effect of attractive interactions on the spatial coupling between translational and rotational facilitation. From the height of the first peak of the functions defined in Eqns. \eqref{MuFunctions} and \eqref{MixedMu}, we computed the facilitation coupling coefficient 
\begin{equation}
C_{F}(a_r,a_{\theta},t_m) = \frac{\mu_{r\theta}^{max}(t_m)\mu_{\theta r}^{max}(t_m)}{\mu_{rr}^{max}(t_m)\mu_{\theta\theta}^{max}(t_m)}
\end{equation}
where $t_m = \text{max}(t_r,t_{\theta})$. $C_{F}(a_r,a_{\theta},t_m)$ is greater than 1 if excitations in one degree of freedom have a greater influence on relaxation in the other degree of freedom as compared to the same one. We see that $C_{F}(a_r,a_{\theta},t_m)$ is lower for $\Delta u/k_BT =$ 1.16 as compared to the repulsive case ($\Delta u/k_BT =$ 0), and increases upon further increasing $\Delta u/k_BT$, especially at large $\phi$ (Fig. \ref{Fig3}A). Next, we examined the rotational and translational contributions to $C_{F}(a_r,a_{\theta},t_m)$ separately, by defining the coefficients $C_{F}^{\theta} = \mu_{\theta r}^{max}/\mu_{\theta\theta}^{max}$ and $C_{F}^{r} = \mu_{r\theta}^{max}/\mu_{rr}^{max}$. We observe that $C_{F}^{\theta}$ varies little across $\phi$ as well as $\Delta u/k_BT$, and the behaviour of $C_{F}(a_r,a_{\theta},t_m)$ is dominated by $C_{F}^{r}$ (Fig. \ref{Fig3}B). This reinforces the finding of Fig. 2 that translational excitations play a more crucial role in rotational relaxation than vice versa.

It has been shown in simulations \cite{keys2011excitations} as well as experiments \cite{gokhale2014growing} that the hierarchical facilitated dynamics of excitations culminates in string-like cooperative motion \cite{donati1998stringlike} and in particular, excitations are analogous to `microstrings' \cite{gebremichael2004particle}. It is therefore tempting to wonder whether growing dynamic correlations generically emerge from excitation dynamics. If this is indeed the case, the spatial decoupling of rotational and translational facilitation in our system should lead to a decoupling of dynamical heterogeneities (DH) in the two degrees of freedom. To investigate this possibility, we defined a coupling coefficient for DH by identifying the top 10\% of the translationally and rotationally most mobile particles over a time interval $\Delta t$. We first computed the distribution $P(r,\Delta t)$ of the minimum distance between each translationally mobile particle and the set of rotationally mobile particles. We also computed a reference distribution $P^{\ast}(r,\Delta t)$ of the minimum distance of translationally mobile particles from a set of randomly chosen non-mobile particles. We then computed the coupling function 
\begin{equation}
D(\Delta t) = \frac{\int_{0}^{r_{min}}P(r,\Delta t)dr}{\int_{0}^{r_{min}}P^{\ast}(r,\Delta t)dr}
\end{equation}
where $r_{min}$ is the first minimum of the radial pair correlation function. $D(\Delta t)$ is analogous to the mobility transfer function defined in \cite{vogel2004spatially}, except that instead of considering translationally mobile particles in two subsequent intervals, we consider translationally and rotationally mobile particles in the \textit{same} interval. We observe that for all $\phi$s (Fig. \ref{Fig3}C) and depletion interaction strengths $\Delta u/k_BT$ considered (Fig. S2), $D(\Delta t)$ shows a peak around $\Delta t \sim t^{\ast}$, where $t^{\ast}$ is the cage-breaking time i.e. the time at which the non-Gaussian parameter $\alpha_2(t)$ reaches a maximum. We chose the maximum value of $D(\Delta t)$, $D_{\text{max}}$, as the coupling coefficient for DH. We observe that for all $\Delta u/k_BT$, $D_{\text{max}}$ increases with $\phi$, which indicates that the spatial coupling between rotational and translational DH increases on approaching the glass transition (Fig. \ref{Fig3}D). More interestingly, in striking resemblance with the behavior of $C_{F}(a_r,a_{\theta},t_m)$ (Fig. \ref{Fig3}A) we find that for $\Delta u/k_BT =$ 1.16, $D_{\text{max}}$ is lower that the purely repulsive case (Fig. \ref{Fig3}D). On increasing $\Delta u/k_BT$ further, we see that $D_{\text{max}}$ once again begins to increase, which is consistent with the dependence of $C_{F}(a_r,a_{\theta},t_m)$ on $\Delta u/k_BT$ (Fig. \ref{Fig3}A). These findings show that the spatial decoupling of facilitated dynamics indeed results in the spatial decoupling of rotational and translational DH. 

Having established the close link between facilitation and DH in rotational as well as translational dynamics, we examine in the influence of attractive interactions on the concentration of translational and rotational excitations $c_{r}$ and $c_{\theta}$, respectively (Eqn. \ref{ExcitConc}). Within the DF theory, the concentration of excitations must vanish at the glass transition. The $\phi$ dependence of $c_{r}$ and $c_{\theta}$ can therefore provide valuable information on the location of glass transitions in our system. Figure \ref{Fig3}E shows the $\phi$ dependence of the concentration of translational excitations, $c_{r}$, for three different strengths of the attractive interaction $\Delta u/k_BT$. We see that for $\Delta u/k_BT =$ 0, $-ln(c_r)$ appears to diverge at some $\phi = \phi_c^{r}$, which we identify as the translational glass transition. Interestingly, with increasing attraction, $\phi_c^{r}$ first shifts to a larger value for $\Delta u/k_BT =$ 1.16, and then decreases upon further increasing $\Delta u/k_BT$ to 1.47. The same $\Delta u/k_BT$ dependence is also exhibited by the orientational glass transition area fraction $\phi_c^{\theta}$ (Fig. \ref{Fig3}F). Further, we observe that this trend is robust and persists for different values of $a_r$ and $a_{\theta}$ (Fig. S3). This shows that the presence of attractive interactions induces reentrant glass transitions in translational as well as orientational degrees of freedom (Fig. \ref{Fig3}G). The presence of reentrant glass transitions in this system was previously inferred \cite{mishra2013two} from the Mode Coupling Theory (MCT) scaling of relaxation times \cite{schilling1997mode,franosch1997theory}. However, we emphasize that the prediction of the DF theory is based on the statistics of localized dynamical events that occur on timescales far smaller than the structural relaxation time. Further, we note that the $\phi_c^{r}$ and $\phi_c^{\theta}$ values obtained from the MCT scaling analysis \cite{mishra2013two} are smaller than the corresponding values obtained here (Fig. \ref{Fig3}G). This is possibly a reflection of the fact that the MCT scaling yields the laboratory glass transition at $\phi = \phi_g$, whereas the concentration of excitations yields the jamming transition at $\phi = \phi_J$. Finally, we note that the shift in the glass transitions is consistent with the spatial decoupling of facilitated dynamics as well as dynamical heterogeneities. Since $C_{F}(a_r,a_{\theta},t_m)$ and $D_{\text{max}}$ increase with $\phi$ (Fig. \ref{Fig3}A \& D), they are indicators of the system's proximity to the glass transition. The addition of attractive interaction lowers both $C_{F}(a_r,a_{\theta},t_m)$ and $D_{\text{max}}$ for a given $\phi$. This indicates that the system moves further away from the glass transition on introducing attractions, which in turn implies that the glass transition shifts to larger $\phi$, as seen in Figure. \ref{Fig3}G.    

\section*{Conclusions}
In summary, we have expanded the concepts of localized excitations and facilitated dynamics to include systems with orientational degrees of freedom (Fig. \ref{Fig1}). By applying this extended framework of DF to experimental data on suspensions of colloidal ellipsoids, we have shown that translational excitations have a more pronounced influence on orientational dynamics than vice versa (Fig. \ref{Fig2}). Further, we observe that the spatial decoupling of translational and rotational dynamical heterogeneities is a consequence of the spatial decoupling of facilitation in these two degrees of freedom (Fig. \ref{Fig3}A \& D). Most importantly, our results show that the reentrance in translational as well as orientational glass transitions with increasing $\Delta u/k_BT$ (Fig. \ref{Fig3}G) can be predicted from the $\phi$ dependence of the concentration of excitations, sans prior knowledge of relaxation times (Fig. \ref{Fig3}E-F). Collectively, these findings show that dynamical facilitation plays a crucial role in governing structural relaxation even in systems with orientational degrees of freedom. In a broader context, the predictive capability of the facilitation approach demonstrated here strongly suggests that a one to one correspondence between the concentration of excitations and the relaxation time, as envisioned in the DF theory, may indeed exist. Our findings immediately open the door for testing the postulates of the DF theory in a variety of complex colloidal systems \cite{glotzer2007anisotropy}. In particular, it would be instructive to see whether the splitting of rotational and translational glass transitions for ellipsoids with large aspect ratio \cite{letz2000ideal,de2007dynamics,zheng2011glass} can be explained within the context of facilitation. On the theoretical front, it would be fascinating to examine whether the hierarchy of excitation energies that scale logarithmically with excitation size \cite{keys2011excitations} also exists for orientational degrees of freedom. Further, it would be interesting to see whether the concentration of rotational excitations can yield quantitative predictions for the temperature as well as density dependence of the orientational relation time. Another promising avenue is the search for potential connections between structural order and facilitation. In the context of ellipsoids, a link between local order and dynamical heterogeneities has already been forged \cite{mishra2013two,zheng2014structural}. Given the connection between DF and heterogeneities observed here, it would be worthwhile to explore whether the spatial occurrence of excitations is itself dictated by local order. We expect our findings to spur further experimental and theoretical research aimed at answering these questions and thereby serve as a stepping stone to a complete understanding of the glass transition.

\section*{Materials and Methods}
Details of the materials and methods have been described in \cite{mishra2013two}. Briefly, colloidal polystyrene ellipsoids with major axis $2l =$ 2.1 $\mu$m and minor axis $2w =$ 1 $\mu$m were synthesized using the method prescribed in \cite{ho1993preparation}. The polydispersities in the major and minor axes are 11\% and 8\%, respectively. Depletion interactions were introduced by adding the non-adsorbing polymer sodium carboxyl methyl cellulose (NaCMC, Fischer-Scientific, mol. wt. 700000, r$_{g}$ 60 nm). Sampled were loaded in wedge-shaped cells and the area fraction $\phi$ was tuned by controlled sedimentation of the ellipsoids to a monolayer-thick region of cells. The plane containing the axes of the ellipsoids was parallel to the walls of the cell. Samples were imaged using a 100X oil immersion objective (Leica, Plan-Apochromat, NA 1.4) and videos were captured at a frame rate of 5 frames per second for 20 minutes. The center of mass coordinates as well as orientations of the ellipsoids were extracted using ImageJ and trajectories were constructed using standard Matlab particle tracking algorithms \cite{crocker1996methods}. Subsequent analysis was performed using Matlab codes developed in-house.  

\section*{acknowledgments}
S.G. thanks the Council for Scientific and Industrial Research (CSIR), India for a Shyama Prasad Mukherjee Fellowship. K.H.N. thanks CSIR, India for a Senior Research Fellowship. R.G. thanks the International Centre for Materials Science (ICMS) and the Jawaharlal Nehru Centre for Advanced Scientific Research (JNCASR) for financial support and A.K.S. thanks Department of Science and Technology (DST), India for support under J.C. Bose Fellowship.

\section*{Corresponding Author Information}
\noindent Shreyas Gokhale\\
Department of Physics\\
Indian Institute of Science\\
Bangalore - 560012, INDIA\\
Ph. no: +91 80 22932579\\
Email: gokhale@physics.iisc.ernet.in\\

\bibliography{references}
\bibliographystyle{apsrev4-1}

\newpage

\begin{figure}[tbp]
\begin{center}
\includegraphics[width=\textwidth]{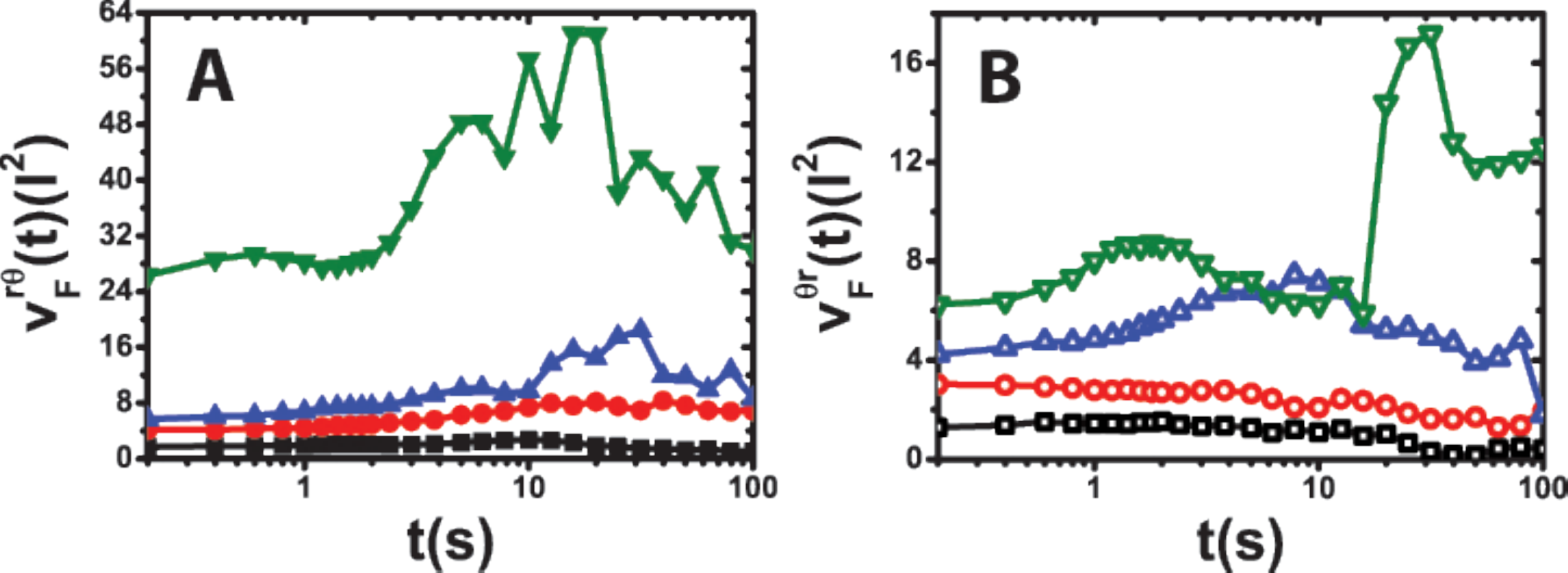}
\end{center}
\noindent \textbf{Fig. S1:} Off-diagonal facilitation volumes for $a_r = 0.5l$ and $a_{\theta} = 15^{\circ}$ (A) $v_{F}^{r\theta}(t)$ for $\phi =$ 0.68 ({\color{black} $\boldsymbol \blacksquare$}), $\phi =$ 0.73 ({\color{red} $\boldsymbol \bullet$}), $\phi =$ 0.76 ({\color{blue} $\boldsymbol \blacktriangle$}) and $\phi =$ 0.79 ({\color{green!50!black} $\boldsymbol \blacktriangledown$}). (B) $v_{F}^{\theta r}(t)$ for $\phi =$ 0.68 ({\color{black} $\boldsymbol \square$}), $\phi =$ 0.73 ({\color{red} $\boldsymbol \circ$}), $\phi =$ 0.76 ({\color{blue} $\boldsymbol \triangle$}) and $\phi =$ 0.79 ({\color{green!50!black} $\boldsymbol \triangledown$}). Note that $v_{F}^{r\theta}(t)$ shows a stronger dependence on $\phi$ as compared to $v_{F}^{\theta r}(t)$.
\label{S1}
\end{figure}

\begin{figure}[tbp]
\begin{center}
\includegraphics[width=\textwidth]{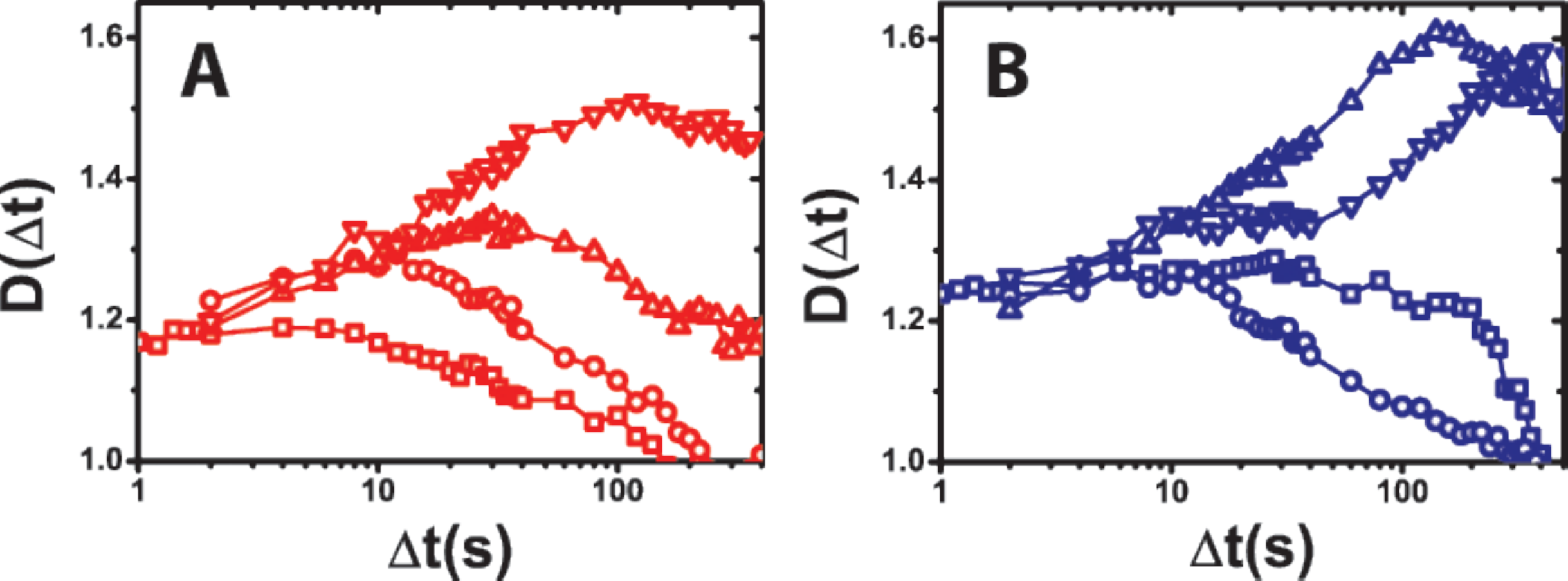}
\end{center}
\noindent \textbf{Fig. S2:} Coupling function for dynamical heterogeneities $D(\Delta t)$. (A) $D(\Delta t)$ for $\Delta u/k_{B}T =$ 1.16 for $\phi =$ 0.67 ({\color{red} $\boldsymbol \square$}), $\phi =$ 0.73 ({\color{red} $\boldsymbol \circ$}), $\phi =$ 0.77 ({\color{red} $\boldsymbol \triangle$}) and $\phi =$ 0.81 ({\color{red} $\boldsymbol \triangledown$}). (B) $D(\Delta t)$ for $\Delta u/k_{B}T =$ 1.47 for $\phi =$ 0.665 ({\color{blue!60!black} $\boldsymbol \square$}), $\phi =$ 0.72 ({\color{blue!60!black} $\boldsymbol \circ$}), $\phi =$ 0.78 ({\color{blue!60!black} $\boldsymbol \triangle$}) and $\phi =$ 0.81 ({\color{blue!60!black} $\boldsymbol \triangledown$}).
\label{S2}
\end{figure}

\begin{figure}[tbp]
\begin{center}
\includegraphics[width=\textwidth]{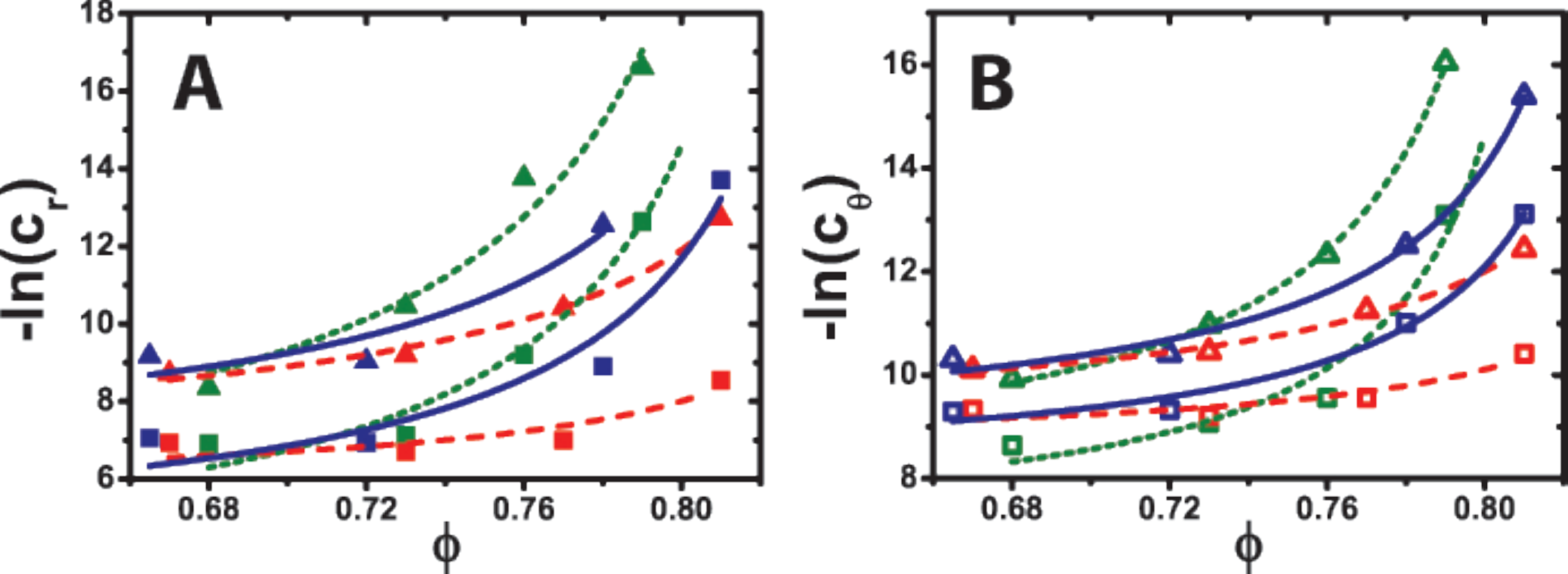}
\end{center}
\noindent \textbf{Fig. S3:} Dependence of $c_r$ and $c_{\theta}$ on $\phi$ (A) Concentration of translational excitations $c_r$ for $a_r = 0.33l$ (filled squares) and $a_r = 0.0.67l$ (filled triangles), for $\Delta u/k_BT =$ 0 (green), $\Delta u/k_BT =$ 1.16 (red) and $\Delta u/k_BT =$ 1.47 (blue). (B) Concentration of  rotational excitations $c_{\theta}$ for $a_{\theta} = 15^{\circ}$ (open squares) and $a_{\theta} = 25^{\circ}$ (open triangles) for $\Delta u/k_BT =$ 0 (green), $\Delta u/k_BT =$ 1.16 (red) and $\Delta u/k_BT =$ 1.47 (blue). In (A) and (B), the curves are empirical fits of the form $\phi_0 + A(\phi_c - \phi)^{-1}$.
\label{S3}
\end{figure}

\newpage
\begin{figure}[tbp]
\includegraphics[width=0.9\textwidth]{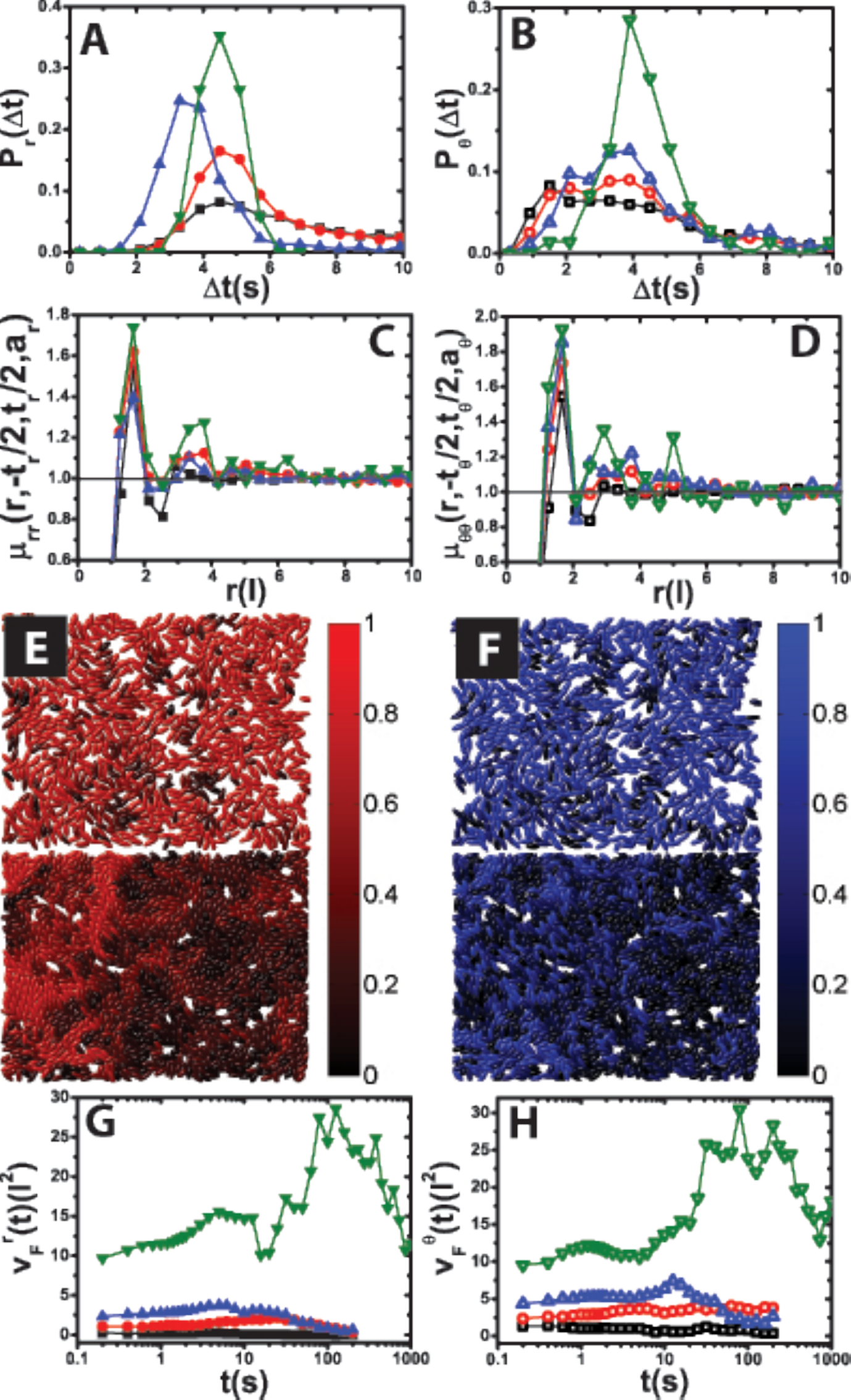}
\end{figure}

\begin{figure}[tbp]
\includegraphics[width=\textwidth]{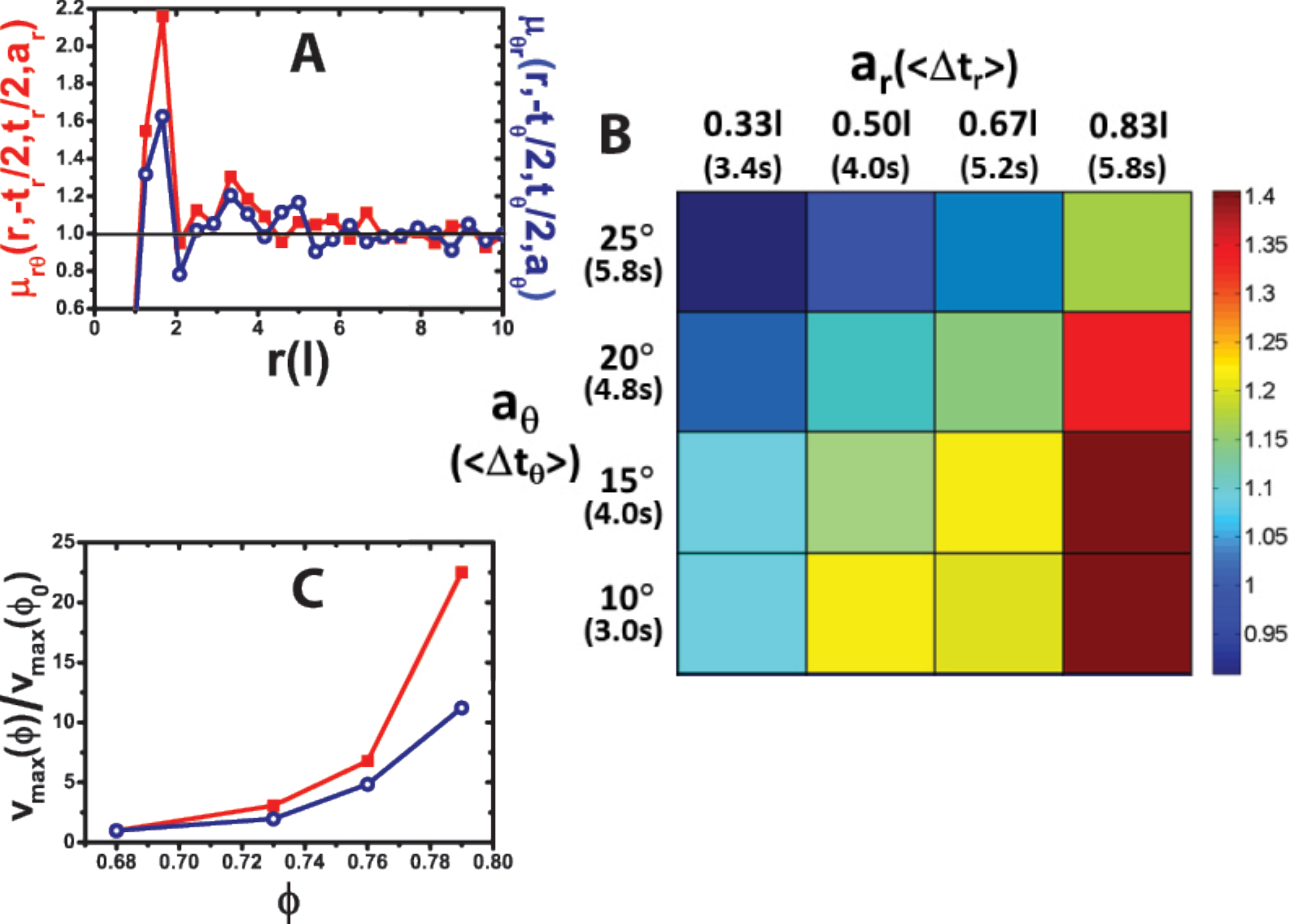}
\end{figure}

\begin{figure}[tbp]
\includegraphics[width=\textwidth]{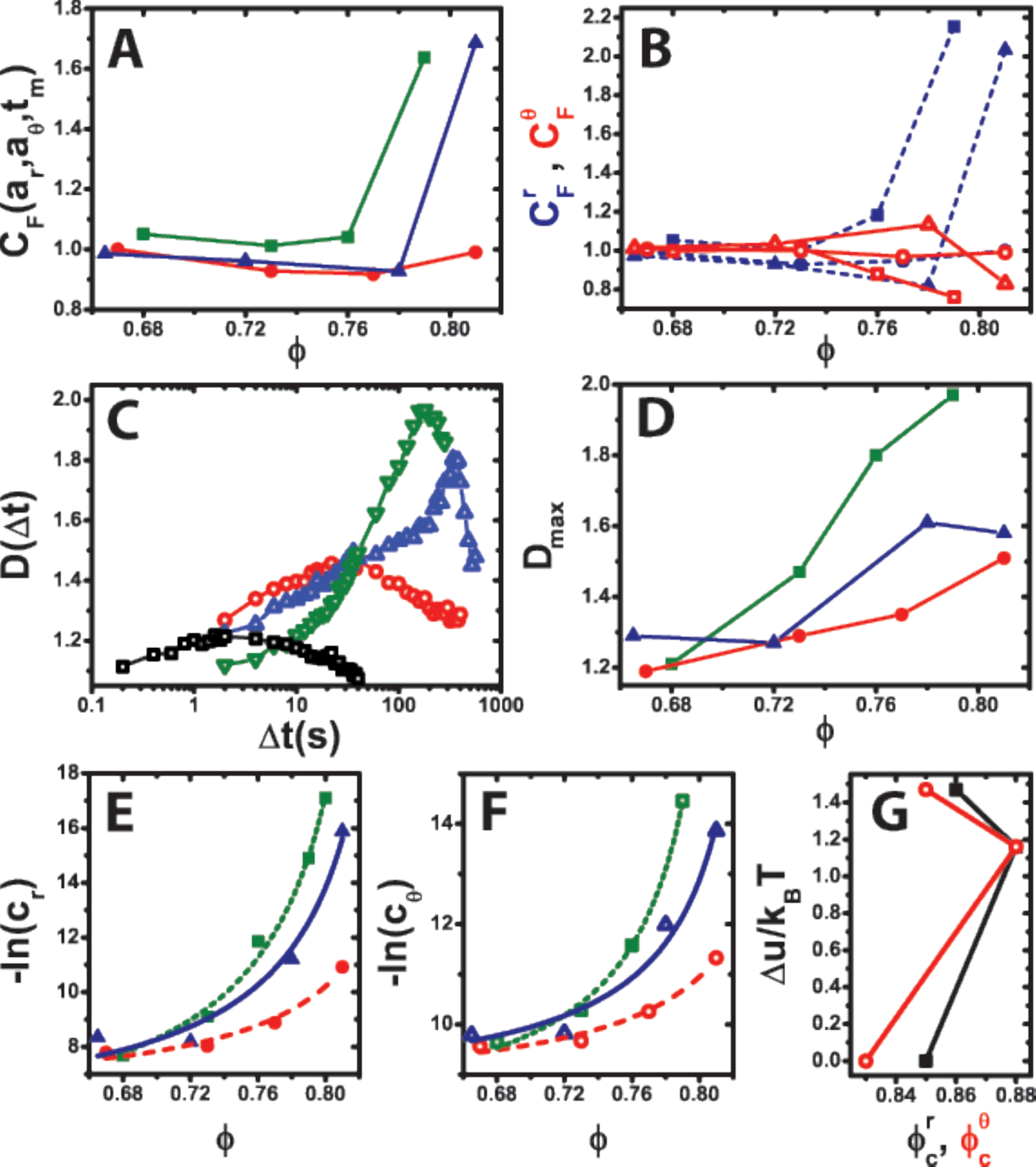}
\end{figure}
\end{document}